\documentclass{JHEP3} 
\conference{Twenty-sixth Johns Hopkins Workshop}                

\usepackage{epsfig,multicol}                    

\newcommand{\lwig}{\mbox{\,\raisebox{.3ex}
    {$<$}$\!\!\!\!\!$\raisebox{-.9ex}{$\sim$}\,}}
\newcommand{\gwig}{\mbox{\,\raisebox{.3ex}
    {$>$}$\!\!\!\!\!$\raisebox{-.9ex}{$\sim$}}\,}
\newbox\mybox
\newcommand\fverb{\setbox\mybox=\hbox\bgroup\verb}
\newcommand\fverbdo{\egroup\medskip\noindent\fbox{\unhbox\mybox}\ }
\newcommand\fverbit{\egroup\item[\fbox{\unhbox\mybox}]}

\newif\ifjhep
\newif\ifhepph

\hepphtrue

\ifhepph\preprint{DESY 03-008\\ hep-ph/0302112}\fi

\title{
       \boldmath
       From QCD Instantons at HERA to Electroweak $B+L$ Violation at 
       VLHC\ifhepph\footnote{Invited talk
       presented at the 26th Johns Hopkins Workshop on Current Problems in Particle Theory, 
       August 1-3, 2002, Heidelberg, Germany.}\fi }

\author{\ifjhep\speaker{\fi Andreas Ringwald\ifjhep}\fi
        \\
        Deutsches Elektronen-Synchrotron DESY, Notkestra\ss e 85, D--22607 Hamburg, Germany\\
        E-mail: \email{andreas.ringwald@desy.de}}

\abstract{
This review emphasizes the close analogy between 
hard QCD in\-stan\-ton-induced chirality 
violating processes in deep-inelastic scattering and 
electroweak in\-stan\-ton-induced baryon plus lepton number ($B+L$) 
violating processes in Quantum Flavor Dynamics (QFD). 
Recent information about QCD instantons, both from lattice simulations and from the 
H1 experiment at HERA, can be used to learn about the fate of  
electroweak $B+L$ violation at future high energy colliders in the 
hundreds of TeV regime, such as the projected 
Very Large Hadron Collider (VLHC). 
The cross-sections turn out to be unobservably small in a conservative
fiducial kinematical region inferred from the above mentioned QCD--QFD analogy. 
An extrapolation -- still compatible with lattice results and 
HERA -- beyond this conservative limit indicates possible observability 
at VLHC.}

\begin{document} 

\section{Introduction}

Over the last decades we have witnessed the remarkable success of 
the Standard Model of electroweak (Quantum Flavor Dynamics (QFD)) and 
strong (QCD) interactions. This success is largely based
on the possibility to apply ordinary perturbation theory to the 
calculation of hard, short-distance dominated scattering processes, since
the relevant gauge couplings are small. 

There are certain processes, 
however, which fundamentally can not be described by ordinary
perturbation theory, no matter how small the gauge coupling is. These processes 
are associated with axial 
anomalies~\cite{Adler:gk
} and 
manifest themselves as anomalous violation of baryon plus lepton  number ($B+L$) in 
QFD and chirality ($Q_5$) in QCD~\cite{'tHooft:up
}. They are induced by topological fluctuations of the non-Abelian gauge
fields, notably by instantons~\cite{Belavin:fg}.

A number of non-perturbative issues in the Standard Model can be understood 
in terms of such topological fluctuations and the associated anomalous processes. 
On the one hand, QCD instantons seem to play 
an important role in various
long-distance aspects of QCD, such as providing a possible solution to 
the axial $U(1)$ problem~\cite{'tHooft:up
} or being at work in
$SU(n_f)$ chiral symmetry breaking~\cite{Shuryak:1981ff,
Schafer:1996wv
}. In QFD, on the other hand, 
similar topological fluctuations of the gauge fields and
the associated $B+L$ violating processes are very important at high 
temperatures~\cite{Kuzmin:1985mm
} and have therefore a crucial impact on the evolution of the baryon and lepton
asymmetries of the universe~\cite{Rubakov:1996vz}.

A related question is whether manifestations of such topological
fluctuations may be directly observed in high-energy scattering processes.
It has been raised originally in the late eighties in 
the context of QFD~\cite{Aoyama:1986ej,Ringwald:1989ee
}. But, despite considerable 
theoretical~\cite{McLerran:1989ab,
Zakharov:1990dj,
Khoze:1991mx,Shuryak:1991pn} and 
phenomenological~\cite{Farrar:1990vb,
Gibbs:1994cw,
Morris:1991bb
} efforts, the actual
size of the cross-sections in the relevant, tens of TeV energy regime was never established (for  
reviews, see Refs.~\cite{Rubakov:1996vz,Mattis:1991bj
}).
Meanwhile, the focus switched to quite analogous QCD instanton-induced hard 
scattering processes in deep-inelastic
scattering~\cite{Balitsky:1993jd,Ringwald:1994kr},  
which are calculable from first principles within instanton-perturbation theory~\cite{Moch:1996bs}, 
yield sizeable rates for observable final state signatures  
in the fiducial regime of the 
latter~\cite{Ringwald:1998ek,Ringwald:1999ze,Carli:1998zf,Ringwald:1999jb,Ringwald:2000gt}, 
and are actively searched for at HERA~\cite{Adloff:2002ph}. 
Moreover, it has been recognized recently that larger-size QCD instantons, beyond the 
semi-classical, instanton-perturbative  regime, may well be responsible
for the bulk of inelastic hadronic processes and build up soft diffractive 
scattering~\cite{Kharzeev:2000ef
}. 

In this review, we emphasize the close analogy of 
QFD and hard QCD instanton-induced processes in deep-inelastic scattering~\cite{Ringwald:1994kr}  
(Sect.~\ref{hard}) and recall the recent 
information about the latter both from lattice 
simulations~\cite{Ringwald:1999ze,Ringwald:2000gt,Smith:1998wt} and from 
the H1 experiment at HE\-RA~\cite{Adloff:2002ph} (Sect.~\ref{HERA}). 
We summarize a recent state of the art evaluation of QFD 
instanton-induced parton-parton cross-sections~\cite{Ringwald:2002sw} (Sect.~\ref{VLHC}), 
as relevant at future high energy colliders in the 
hundreds of TeV regime, such as the projected 
Eurasian Long Intersecting
Storage Ring (ELOISATRON)~\cite{eloisatron} or the 
Very Large Hadron Collider (VLHC)~\cite{vlhc}.  
The implications of the lattice and HERA results -- 
via the above mentioned QFD--QCD analogy -- 
for the fate of electroweak $B+L$ violation in high energy collisions are discussed.  

\section{\label{hard}Instanton-induced hard scattering processes}

\subsection{Generalities}

Instantons~\cite{'tHooft:up,Belavin:fg,
Affleck:1980mp} are minima of the classical Euclidean Yang-Mills action, 
localized in space and Euclidean
time, with unit topological charge (Pontryagin index) $Q=1$. 
In Minkowski space-time, instantons  
describe tunneling transitions between classically degenerate, topologically
inequivalent vacua, differing in their 
winding number (Chern-Simons number) 
by one unit, $\triangle N_{\rm CS}=Q=1$~\cite{Jackiw:1976pf
}. 
The corresponding energy barrier (``sphaleron energy''~\cite{Klinkhamer:1984di}), 
under which the instantons tunnel, is inversely proportional to $\alpha_g\equiv g^2/(4\pi )$, 
the fine-structure constant of the relevant gauge theory,  
and the effective instanton-size $\rho_{\rm eff}$, 
\begin{eqnarray}
\label{barrier}
   M_{\rm sp}\sim \frac{\pi}{\alpha_g\,\rho_{\rm eff}}    \sim
               \cases { \pi \frac{M_W}{\alpha_W}\sim 10\ {\rm TeV} &{\rm in\ QFD~\cite{Klinkhamer:1984di}}\,,\cr
                        {\mathcal Q} & 
{\rm in\ QCD~\cite{Ringwald:1994kr,Moch:1996bs,Ringwald:1998ek}}\,, \cr}
\end{eqnarray}
where ${\mathcal Q}$ is a large 
momentum transfer e.g. in deep-inelastic scattering (DIS), which should be taken $\gwig 10$ GeV in order
to be in the semi-classical, 
instanton-perturbative regime~\cite{Moch:1996bs,Ringwald:1998ek,Ringwald:1999ze,Ringwald:2000gt}.
Axial anomalies~\cite{Adler:gk
} force instanton-induced hard scattering processes 
to be always associated with anomalous fermion-number violation~~\cite{'tHooft:up
}, in particular $B+L$ 
violation, 
\begin{equation}
\triangle  B = \triangle L =-n_{\rm gen}\, Q ,
\end{equation} 
in the case of QFD  with $n_{\rm gen}=3$ fermion generations, 
and chirality violation, 
\begin{equation}
\triangle Q_5 = 2\,n_f\,Q ,
\end{equation} 
in the case of QCD with typically $n_f=3$ light quark flavors.   

\FIGURE{\epsfig{file=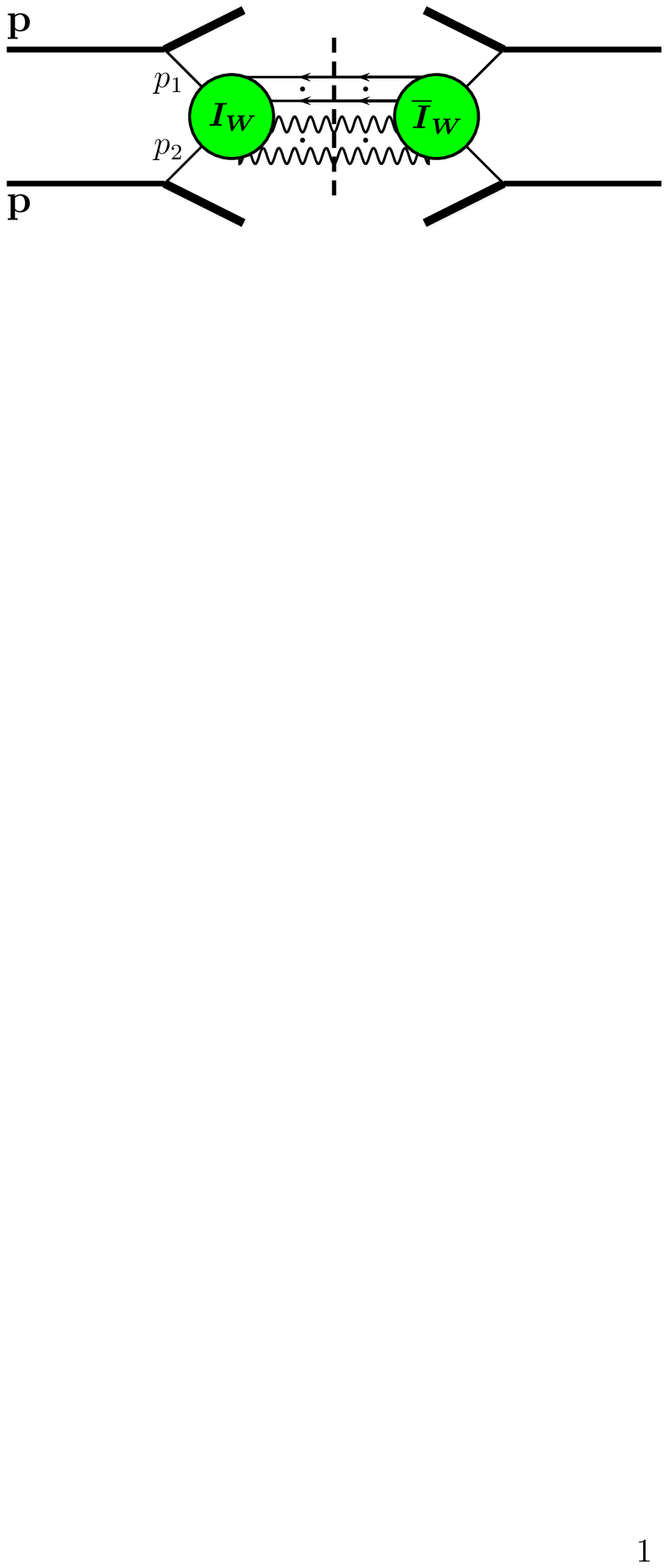,width=6.5cm,bbllx=74pt,bblly=619pt,bburx=314pt,bbury=714,clip=}%
\hspace{6ex}
        \epsfig{file=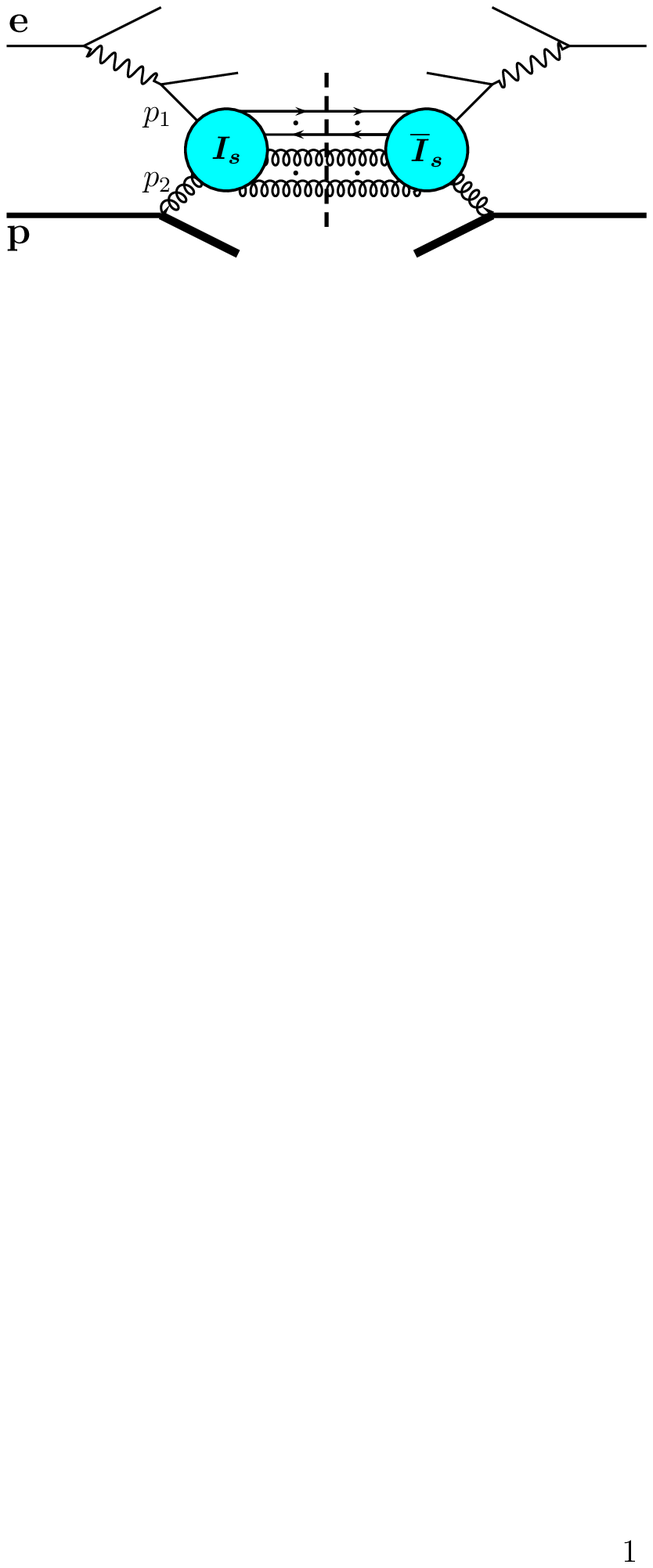,width=6.5cm,bbllx=74pt,bblly=619pt,bburx=314pt,bbury=714,clip=}%
        \caption{Illustration of a QFD instanton-induced process in proton-proton scattering ({\em left}) and of a 
QCD instanton-induced process in deep-inelastic electron-proton scattering ({\em right}) 
 (from~\cite{Ringwald:2002sw}).}%
        \label{opt_illu}}

In instanton-perturbation theory -- the semi-classical expansion
of the corresponding path integral expressions about the instanton solution --  
instanton-induced total cross-sections for hard parton-parton ($\rm p_1$-$\rm p_2$) 
scattering processes
(cf. Fig.~\ref{opt_illu}) are given in terms of an integral over the 
instanton-anti-instanton\footnote{\label{iai}Both an
instanton and an anti-instanton enter here, 
since the cross-section~(\ref{gencross}) has been written as a discontinuity of the $\rm p_1\,p_2$ 
forward elastic scattering amplitude in the
$I\bar I$-background (cf. Fig.~\ref{opt_illu}). 
Alternatively, one may calculate the cross-section by taking the modulus squared of 
amplitudes in the single instanton-background.} ($I\bar I$) collective coordinates 
(sizes $\rho,\bar\rho$, $I\bar I$ distance $R$, relative color 
orientation $U$)~\cite{Ringwald:1998ek} (see also~\cite{Zakharov:1990dj,
Khoze:1991mx,Shuryak:1991pn,Ringwald:2002sw,Klinkhamer:1991pq,
Balitsky:1992vs})
\begin{eqnarray}
\label{gencross}
      \hat\sigma^{(I)}_{\rm p_1p_2} &= &
\frac{1}{2\,p_1\cdot p_2}\,{\rm Im}\,
      \int {\rm d}^4 R\ 
   {\rm e}^{{\rm i}\, ( p_1+p_2 )\cdot R}
\\[1.5ex] \nonumber \mbox{} && \times\ 
      \int\limits_0^\infty {\rm d}\rho 
      \int\limits_0^\infty {\rm d}\overline{\rho}\  
      { D(\rho)\, D(\overline{\rho})}
       \int {\rm d}U\ 
            {\rm e}^{{-\frac{4\pi}{\alpha_g 
                                            }}
      { \Omega\left(U, \frac{R^2}{\rho\bar\rho},\frac{\bar\rho}{\rho}, \ldots \right)}}
\ 
      \left[\omega\left(U, \frac{R^2}{\rho\bar\rho},\frac{\bar\rho}{\rho},\ldots \right)\right]^{n_{\rm fin}}  
\\[1.5ex] \nonumber \mbox{} && \times\  
F\left(\sqrt{-p_1^{2}}\,\rho\right)\,
      F\left(\sqrt{-p_1^{2}}\,\overline{\rho}\right)\,
      F\left(\sqrt{-p_2^2}\,\rho\right)\,F\left(\sqrt{-p_2^2}\,\overline{\rho}\right)
\,
\ldots 
\,. 
\end{eqnarray}
Here, the main building blocks are {\em i)} the instanton-size distribution $D(\rho )$, {\em ii)} the
function $\Omega$, which takes into account the exponentiation of the production of 
${\mathcal O}(1/\alpha_g)$ (gauge) bosons~\cite{Ringwald:1989ee,
Zakharov:1990dj
} and can be identified with the 
$I\bar I$-interaction defined via the valley method~\cite{Khoze:1991mx,Yung:1987zp,Verbaarschot:1991sq,%
Arnold:1991dv
}, 
and {\em iii)} the function $\omega$, whose 
explicit form can be found in Refs.~\cite{Shuryak:1991pn,Ringwald:1998ek} and which summarizes the effects of 
final-state fermions. Their
number $n_{\rm fin}$ is related to the number $n_{\rm in}$ of initial-state fermions via the anomaly,  
\begin{eqnarray}
\label{nfin}
   n_{\rm fin} + n_{\rm in} \equiv  n_{\rm tot} \equiv  
               \cases { 4\,n_{\rm gen}\   = 12  &{\rm in\ QFD}\,,\cr
                        2\,n_f\       &{\rm in\ QCD}\,. \cr}
\end{eqnarray}
With each initial-state parton $p$, there is an associated ``form
factor''~\cite{Moch:1996bs,Ringwald:1998ek}, 
\begin{equation}
\label{formfac}
F(x)=x\,K_1(x)\ \left\{
\begin{array}{lcllcll}
&\simeq & \sqrt{\pi/(2\,x)}\exp(-x) &{\rm \ for\ }&
x& \gg 1 ,\nonumber\\[2mm]
&=& 1 &{\rm \ for\ }& x&=0\,.
\nonumber\\
\end{array}\right.
\end{equation} 
The dots in Eq.~(\ref{gencross}) stand for further known smooth 
factors~\cite{Ringwald:1998ek}.  

Let us elaborate further on the quantities mentioned under {\em i)} and {\em ii)}. 

{\em Ad i)} The instanton-size distribution $D(\rho )$ is known in 
instanton-perturbation theory, $\alpha_g (\rho^{-1} )\ll 1$,
up to two-loop renormalization group invariance~\cite{'tHooft:up,
Bernard:1979qt,Morris:zi}. In QCD, the loop corrections are sizeable in the phenomenologically interesting
range~\cite{Ringwald:1998ek,Ringwald:1999ze}. For a qualitative discussion, however, 
the one-loop expression for the size distribution,    
\begin{eqnarray}
\label{size-dist}
     { D({\rho})} =
     \frac{d}{\rho^5}
      \left(\frac{2\pi}{\alpha_g (\mu )}\right)^{2\,N_c} 
      (\mu\,\rho )^{\beta_0
      } \ 
     {\rm e}^{-\frac{2\pi}{\alpha_g (\mu )}\,{ S^{(I)}}} \,,    
\end{eqnarray} 
suffices, which, moreover, is numerically adequate for the case of QFD because of its weak coupling, 
$\alpha_W(M_W)\equiv \alpha (M_W)/\sin^2 \hat\theta (M_W)=0.033819(23)$~\cite{Hagiwara:pw}.
In Eq.~(\ref{size-dist}), 
\begin{eqnarray}
\label{beta0}
\beta_0 = 
\frac{11}{3}\,N_c - \frac{1}{6}\,n_s - \frac{1}{3}\,n_{\rm tot} = 
               \cases { 19/6   &{\rm in\ QFD} ($N_c=2, n_s=1, n_{\rm tot}=12$)\,,\cr
                        11 - 2\,n_f/3      &{\rm in\ QCD} ($N_c=3, n_s=0, n_{\rm tot}=2\,n_f$)\,, \cr}
\end{eqnarray} 
denotes the first coefficient in the $\beta$ function, 
\begin{eqnarray}
\label{action}
      { S^{(I)}} =
               \cases {1 + \frac{1}{2}\,M_W^2\,\rho^2 
         +{\mathcal O}\left(M_W^4\,\rho^4 \ln (M_W\,\rho) \right)& 
                     {\rm in\ QFD~\cite{'tHooft:up,
Affleck:1980mp}}\,,\cr
                        1 &{\rm in\ QCD~\cite{Belavin:fg}}\,, \cr}
\end{eqnarray}
the instanton action, $\mu$ the renormalization scale, and $d$ a scheme-dependent constant, which
reads in the $\overline{\rm MS}$ scheme~\cite{Hasenfratz:1981tw
},  
\begin{eqnarray}
\label{dmsbar}
d_{\overline{\rm MS}} = 
\frac{2\,{\rm e}^{5/6}}{\pi^2\,(N_c-1)!(N_c-2)!}\,{\rm
 e}^{-1.511374\,N_c+0.291746\,(n_{\rm tot}+n_s)/2}
\,.
\end{eqnarray}  
\FIGURE{\epsfig{file=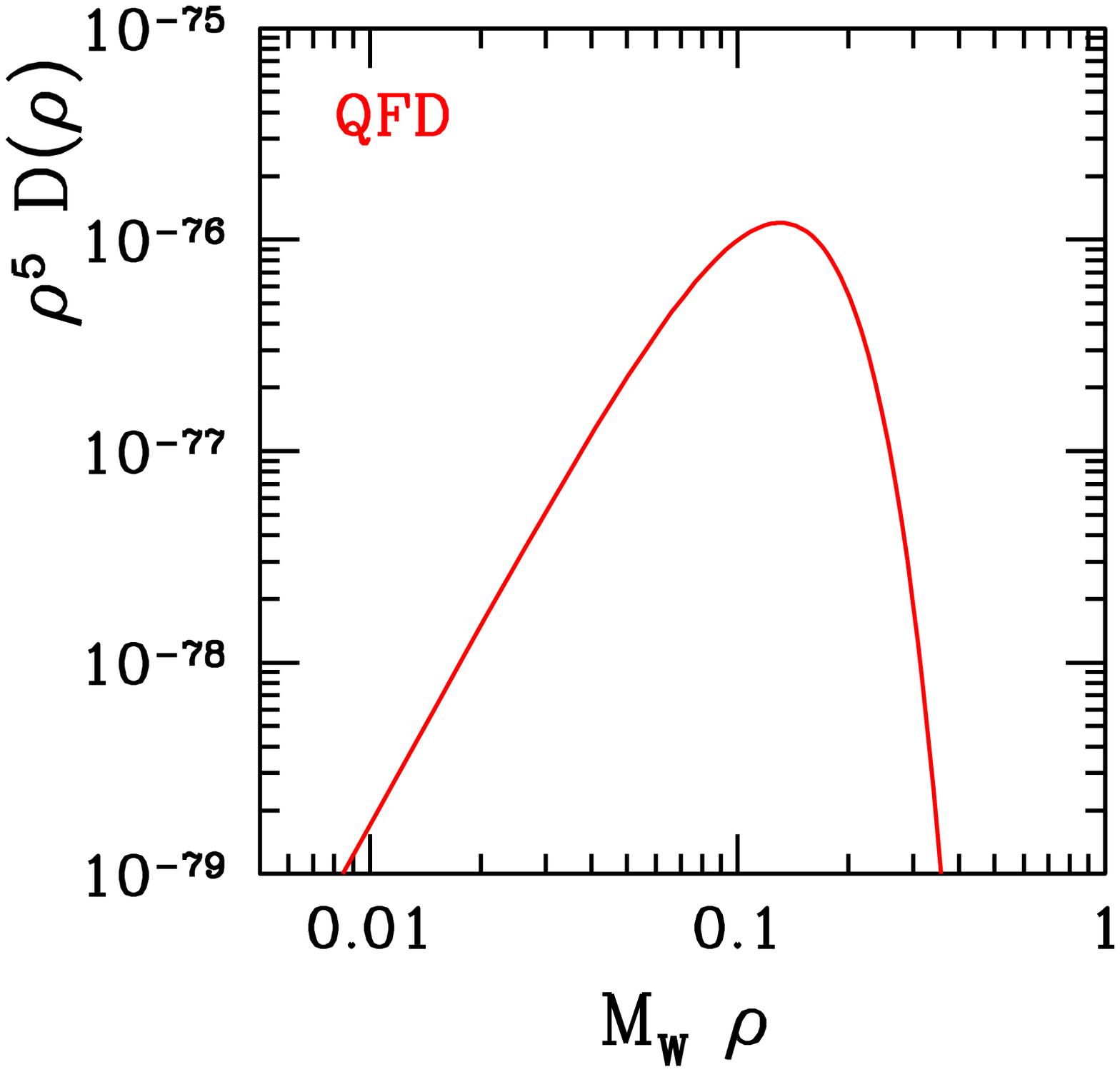,width=7.4cm,clip=}%
\hspace{1ex}
        \epsfig{file=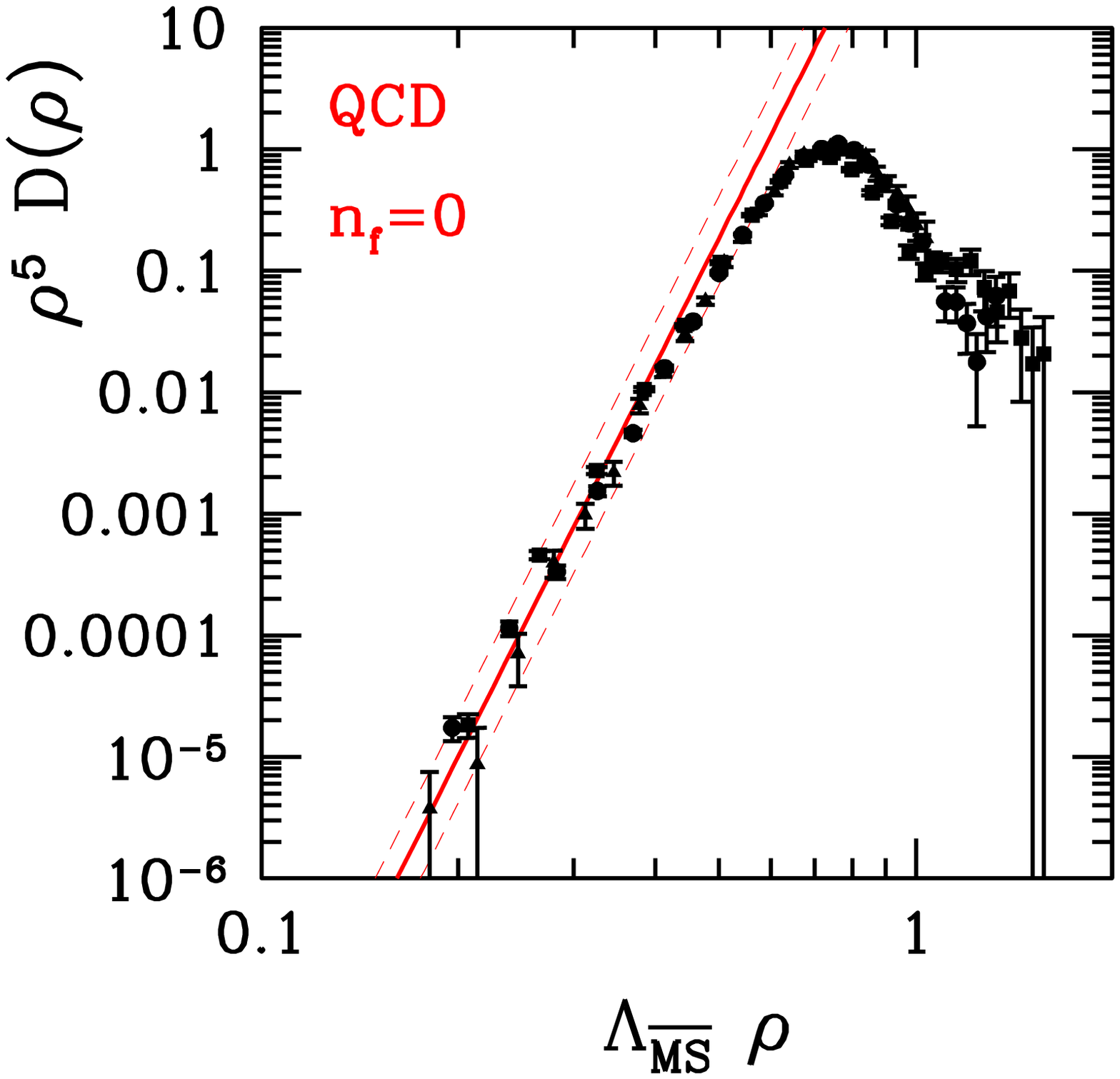,width=7.4cm,clip=}%
        \caption{Instanton-size distributions as predicted in instanton-perturbation theory (solid lines) 
in QFD (left) and quenched ($n_f=0$) QCD (right). 
For QCD (right), the two-loop renormalization group invariant prediction for the size distribution from 
Ref.~\cite{Morris:zi} together with the 3-loop form of $\alpha_{\overline{\rm MS}}$, with
$\Lambda^{(0)}_{\overline{\rm MS}}=238\pm 19$~MeV from the ALPHA collaboration~\cite{Capitani:1998mq}, was used.
The error band (dashed lines) results from the errors in $\Lambda_{\overline{\rm MS}}$ and a variation
of $\mu = 1\div 10$~GeV. 
{\em Left:} Towards large sizes, 
the QFD instanton size distribution decreases
exponentially due to the Higgs mechanism.
{\em Right:} 
For large sizes, 
the QCD instanton size 
distribution, as determined from recent high-quality lattice data from UKQCD~\cite{Smith:1998wt}$^{{\ref{lattice-ref}}}$, 
appears to decrease exponentially, $\propto\exp (-c\rho^2)$~\cite{Ringwald:1999ze,Shuryak:1999fe}, 
similar to the QFD size distribution (left), but unlike the instanton-perturbative prediction (solid).  
For small sizes, 
on the other hand, one observes
a remarkable agreement with the predictions from  
instanton-perturbation theory (solid)~\cite{Ringwald:1999ze,Ringwald:2000gt}. 
}%
        \label{size-dist-illu}}

The quite differenct $\rho$ dependence of 
the size distribution~(\ref{size-dist}) for QFD and QCD has important consequences 
for the predictivity of  
instanton-induced subprocess cross-sections~(\ref{gencross}). 
The validity of instanton-perturbation theory requires instantons of small enough size,
such that $\alpha_g (\rho^{-1})\ll 1$. 
In QFD, this is guaranteed by the exponential decrease 
$\propto\exp (-\pi\,M_W^2\,\rho^2/\alpha_W )$ (cf.~(\ref{action})) 
of the size distribution~(\ref{size-dist}) for 
$\rho > \rho_{\rm max} =0.13/M_W$ (cf. Fig.~\ref{size-dist-illu} (left)). 
Therefore, the relevant contributions to the
size integrals in~(\ref{gencross}) arise consistently from the perturbative region 
($\alpha_W(\rho^{-1})\ll 1$) even if both initial partons are on-shell, $p_1^2\approx p_2^2\approx 0$, as 
relevant for electroweak instanton-induced processes, e.g. in proton-proton 
scattering (cf. Fig.~\ref{opt_illu} (left)).

In QCD, on the other hand, the perturbative expression~(\ref{size-dist}) for the size distribution
behaves power-like, $\propto \rho^{\beta_0 -5}$ (cf. Fig.~\ref{size-dist-illu} (right)). This behavior 
generically causes the dominant contributions to observables like the cross-section~(\ref{gencross}) 
to originate from large $\rho\sim \Lambda^{-1}\Rightarrow \alpha_s(\rho^{-1})\sim 1$ and thus
tends to spoil the applicability of instanton-perturbation theory.
Deep-inelastic scattering, however, offers a unique opportunity to test  
the predictions of instanton-perturbation 
theory from first 
principles~\cite{Moch:1996bs,Ringwald:1998ek,Ringwald:1999ze,Carli:1998zf,Ringwald:1999jb,Ringwald:2000gt}. 
This can be understood as follows.   
In deep-inelastic electron-proton scattering, the virtual photon splits into a quark and an 
antiquark, one of which, $\rm p_1$ say, enters the instanton subprocess (cf. Fig.~\ref{opt_illu} (right)).
This parton carries a space-like virtuality $Q^{\prime\, 2}\equiv -p_1^2\geq 0$, which can be
made sufficiently large by kinematical cuts on the final state. In this hard-scattering regime the 
contribution of large instantons to the integrals  is suppressed by the exponential form 
factors~(\ref{formfac}) in (\ref{gencross}), $\propto {\rm e}^{-Q^\prime (\rho+\bar\rho )}$,    
and instanton-perturbation theory is reliable~\cite{Moch:1996bs,Ringwald:1998ek}. 
In this connection it is quite welcome that 
lattice data on the instanton content of the quenched ($n_f=0$) 
QCD vacuum~\cite{Smith:1998wt}\footnote{\label{lattice-ref}For further, qualitative similar lattice data, 
see Refs.~\cite{Hasenfratz:1998qk,GarciaPerez:1998ru} and the 
reviews~\cite{Negele:1998ev,
Stamatescu:2000ch}.}
can be used to infer the region of validity of instanton-perturbation
theory for $D(\rho )$~\cite{Ringwald:1999ze,Ringwald:2000gt}: 
As illustrated in Fig.~\ref{size-dist-illu} (right), there is very good agreement 
for $\Lambda_{\overline{\rm MS}}\,\rho\lwig 0.42$. 

{\em Ad ii)} A further basic building block of instanton-induced  cross-sections~(\ref{gencross}) is 
the function
$\Omega(U,R^2/(\rho\overline{\rho}),\overline{\rho}/\rho)$, appearing
in the exponent with a large numerical 
coefficient $4\pi/\alpha_g$. It summarizes the effects of the ${\mathcal O}(1/\alpha_g)$ 
final-state (gauge) bosons, mainly $W$'s and $Z$'s in the case of QFD and gluons in the case of 
QCD. Within strict instanton-perturbation theory, it is given in form
of a perturbative expansion~\cite{Zakharov:1990dj,
Ringwald:1998ek,Arnold:1991dv
} for large $I\bar I$-distance squared $R^2$. 
Beyond this expansion, one may identify $\Omega$ with the interaction between an instanton and 
an anti-instanton,  
which may be systematically evaluated by means of  
the so-called $I\bar I$-valley method~\cite{Yung:1987zp}. 
The corresponding interaction has been found analytically for the case of 
pure $SU(2)$ gauge theory\footnote{\label{emb}For the embedding 
of the $SU(2)$ $I\bar I$-valley into $SU(3)$, 
see e.g. Ref.~\cite{Ringwald:1999ze}.}~\cite{Khoze:1991mx,Verbaarschot:1991sq},   
\begin{equation}
\Omega_g
=
\Omega_0 + \Omega_1\,u_0^2 +\Omega_2\,u_0^4\,,
\label{valley}
\end{equation}
with
\begin{eqnarray}
\nonumber
\Omega_0 &=& 2\,\frac{z^4-2z^2+1+2\, (1-z^2)\ln z}{(z^2-1)^3}\,,
\\[1.5ex] \label{valley-int}
\Omega_1 &=& -8\,\frac{z^4-z^2+(1-3\,z^2)\ln z}{(z^2-1)^3}\,,
\\[1.5ex] \nonumber
\Omega_2 &=& -16\,\frac{z^2-1-(1+z^2)\ln z}{(z^2-1)^3}\,.
\end{eqnarray}
Due to conformal invariance of classical pure Yang-Mills theory, it depends on the 
sizes $\rho$, $\bar\rho$, and the $I\bar I$-distance $R$ only through the 
``conformal separation'', 
\begin{equation}
\label{conf-sep}
z = \frac{1}{2} \left( \xi+\sqrt {{\xi}^{2}-4}\right)\,, \hspace{9ex}
\xi = \frac{R^2}{\rho\overline{\rho}}+\frac{\overline{\rho}}{\rho}+
\frac{\rho}{\overline{\rho}}\,\geq\, 2\,, 
\end{equation} 
and on the relative color orientation$^{\ref{emb}}$ $U=u_0+{\rm i}\,\sigma^k u_k$, with $u_0^2+u^k u_k=1$,  
only through $u_0$. 
\FIGURE[t]{
\epsfig{file=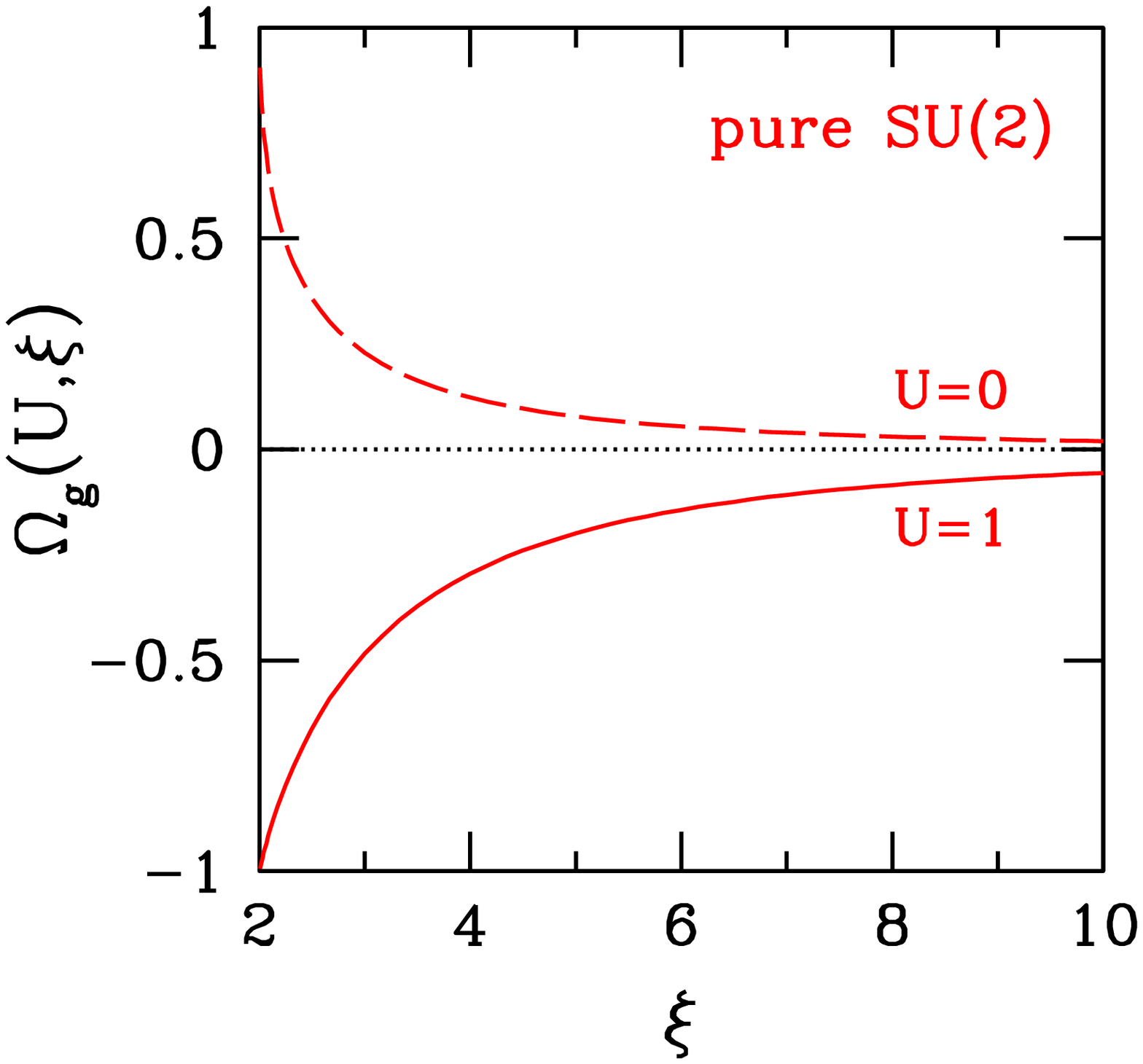,width=7.6cm,clip=}%
\hspace{2ex}
        \epsfig{file=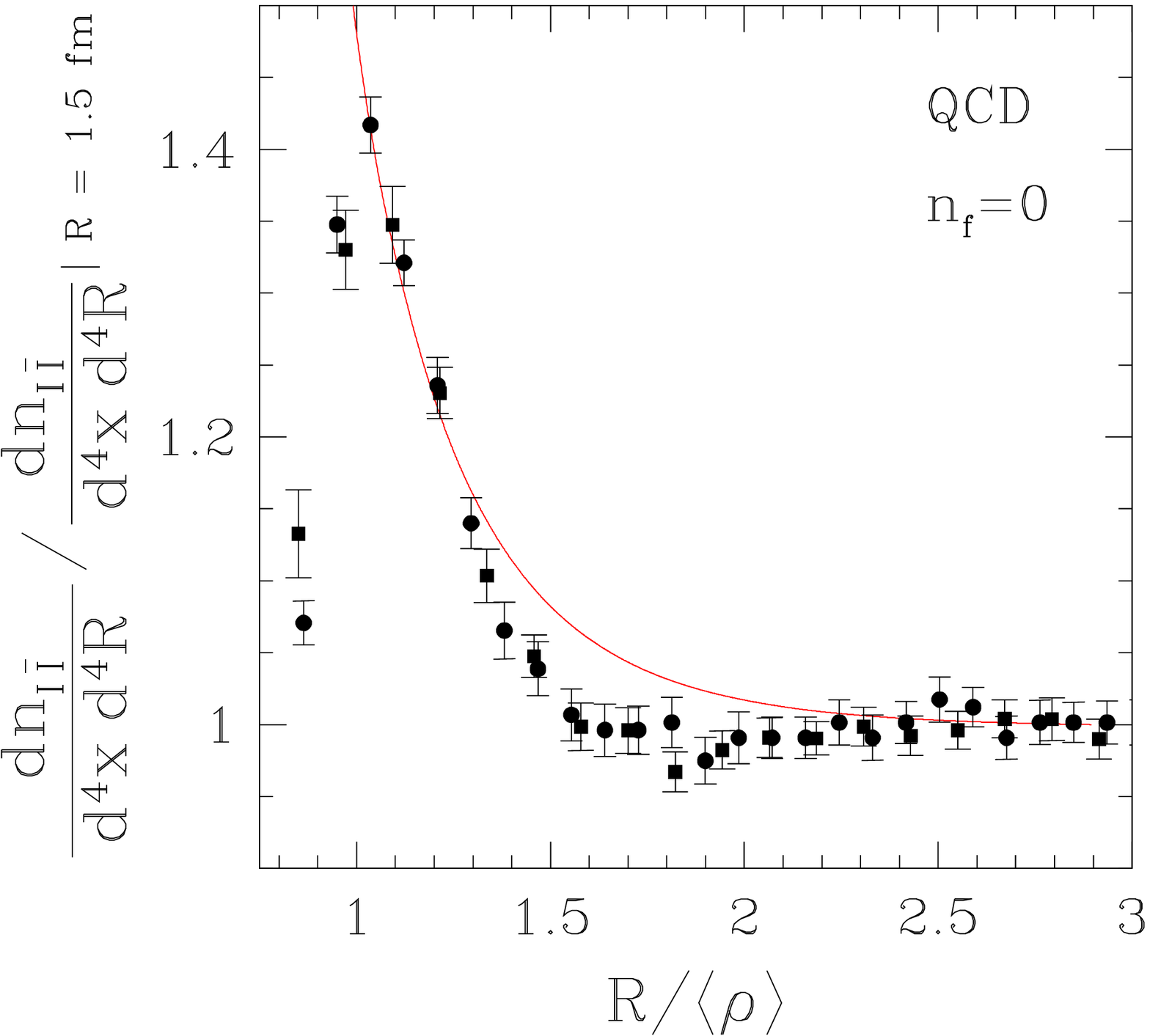,width=6.4cm}%
        \caption{
{\em Left:} $I\bar I$-valley interaction~(\ref{valley}) as function of 
conformal separation $\xi$~(\ref{conf-sep}) for the most attractive 
relative orientation ($U=1$, solid) and the most repulsive relative orientation 
($U=0$, dashed).
{\em Right:}  
Illustration of the agreement
of recent high-quality lattice data~\cite{Smith:1998wt} 
for the $I\overline{I}$-distance
distribution with the predictions from  
instanton-perturbation theory (solid) 
for $R/\rho\geq 1.05$~\cite{Ringwald:1999ze,Ringwald:2000gt}.
}%
   \label{qcd-lattice-dist}}

In the weak coupling regime, $\alpha_g\ll 1$,  
$I\bar I$-pairs with the most attractive relative orientation,  $U=1$,  
give the dominant contribution to the cross-section~(\ref{gencross}).  For this relative 
orientation, the $I\bar I$-valley represents 
a gauge field configuration of steepest descent interpolating between an 
infinitely separated $I\bar I$-pair, corresponding to twice the instanton action, 
$S^{(I\bar I)}=2\,[1+\Omega_g(U=1,\xi=\infty)]=2$, and a strongly overlapping one, 
annihilating to the perturbative vacuum at $\xi = 2$ ($R=0,\rho=\bar\rho$), 
corresponding to vanishing action $S^{(I\bar I)}=2\,[1+\Omega_g(U=1,\xi=2)]=0$ 
(cf. Fig.~\ref{qcd-lattice-dist} (left)).
Thus, near $\xi\approx 2$, the semi-classical approximation 
based on the $I\bar I$-valley breaks down and no reliable non-perturbative information can be extracted 
from it.

In this connection one may exploit again high-quality lattice data~\cite{Smith:1998wt} on 
the $I\bar I$-distance distribution 
in quenched QCD to estimate the fiducial region in  $\xi$   
or more specifically in $R/\langle\rho\rangle$, where $\langle \rho\rangle\approx 0.5$~fm 
is the average instanton/anti-instanton size measured on the lattice (cf. Fig.~\ref{size-dist-illu} (right)).    
Good agreement with the predictions from instanton-perturbation theory 
is found 
for 
$R/\langle\rho\rangle\gwig 1.0\div 1.05$~\cite{Ringwald:1999ze,Ringwald:2000gt} 
(cf. Fig.~\ref{qcd-lattice-dist} (right)). 
There are, however,  remaining ambiguities in this case. 
{\em a)} The integrations over $\rho$, $\bar\rho$ 
in the $I\bar I$-distance distribution ${\rm d}n_{I\bar I}/({\rm d}^4x\, {\rm d}^4 R  )$ imply
significant contributions also from larger instantons with 
$0.42\,\lwig\, \Lambda_{\overline{\rm MS}}\,\rho, \Lambda_{\overline{\rm MS}}\,\overline{\rho}\,\lwig\, 1$, 
outside the region of instanton-perturbation theory. 
A more differential lattice measurement of the distance distribution, 
${\rm d}n_{I\bar I}/({\rm d}^4x\, {\rm d}^4 R\, {\rm d}\rho\, {\rm d}\bar\rho )$,
which includes also differentials with respect to the sizes $\rho$ and $\bar\rho$, 
and eventually a test of its conformal properties would resolve these theoretical ambiguities.   
{\em b)} 
At small $I\bar I$-separation $R<(\rho +\bar\rho)/2$, the extraction of the 
$I\bar I$-distance distribution from the quenched QCD lattice data is quite ambiguous since there is no 
principal distinction between a trivial gauge field fluctuation and an $I\bar I$-pair at small 
separation. 
This is reflected in a considerable dependence on the cooling method/amount used to 
infer properties of the $I\bar I$-distance distribution~\cite{GarciaPerez:1998ru,Stamatescu:2000ch}. 
A simple extrapolation of lattice results on the topological structure of 
quenched $SU(2)$ gauge theory~\cite{GarciaPerez:1998ru} to zero ``cooling radius'' indicates  
$\langle R/(\rho+\bar\rho)/2\rangle\approx 0.5$, i.e. strongly overlapping $I\bar I$-pairs in the vacuum, 
unlike Fig.~\ref{qcd-lattice-dist} (right).   
Therefore, the fiducial region $R^2/(\rho\bar\rho)\geq 1$ for the reliability of instanton-perturbation
theory inferred from lattice data should be considered as quite conservative.

\subsection{Saddle-point evaluation}

For  weak-coupling, $\alpha_g\ll 1$, the collective coordinate integrals in 
the cross-section~(\ref{gencross}) can be performed in the saddle-point approximation, 
the relevant effective exponent being\footnote{\label{qcd-mod}In the case of QCD, 
some additional terms, which 
arise from the running of $\alpha_s$ and are formally of pre-exponential nature, have to be 
included in 
Eqs.~(\ref{holy-exp}) and (\ref{saddle-coll}) for 
numerical accuracy~\cite{Ringwald:1998ek}. The simplified 
expressions~(\ref{holy-exp}),  
(\ref{saddle-coll}) 
suffice, however, for illustrative purposes and are numerically adequate for QFD.}
\begin{eqnarray}
\label{holy-exp}
\lefteqn{
-\Gamma \equiv  {\rm i}\, ( p_1+p_2 )\cdot R}
\\[1.5ex] \nonumber  && -   
               \cases {  
   \frac{4\pi}{\alpha_W(\mu )}
       \left[ 1 + \frac{1}{4}\, M_W^2 (\rho^2+\bar\rho^2 )
       + \Omega_g \left(U, \frac{R^2}{\rho\bar\rho}, \frac{\bar\rho}{\rho}  \right) \right] 
     &[{\rm QFD} ($p_1^2=p_2^2=0$)],\cr
   Q^\prime\,(\rho +\bar\rho ) + \frac{4\pi}{\alpha_s(\mu )}
       \left[ 
       1
       + \Omega_g \left(U, \frac{R^2}{\rho\bar\rho}, \frac{\bar\rho}{\rho}  \right) \right]
      &[{\rm QCD} (DIS: $-p_1^2=Q^{\prime 2}>0, p_2^2=0$)]. \cr}
\end{eqnarray}       
For the case of QFD,  the Higgs part $\Omega_h$ of the 
$I\bar I$-interaction has been  neglected in~(\ref{holy-exp}) and for the gauge part the one 
from the pure gauge theory, $\Omega_g$, was taken.    
This should be considered as reliable as long as the dominant contribution to the QFD instanton-induced cross-section
is due to the multiple production of ${\mathcal O}(1/\alpha_g)$ transverse $W$'s and $Z$'s -- as is the case 
at energies below the sphaleron~(\ref{barrier}) -- rather than of longitudinal ones and 
of Higgs bosons~\cite{Khoze:1991mx}.  

The saddle-point equations, $\partial\Gamma/\partial\chi_{\mid \chi_\ast}=0$, 
with $\chi = \{ U,R,\rho,\bar\rho\}$, 
following from~(\ref{holy-exp}) imply that the instantons and anti-instantons contributing to 
the cross-section~(\ref{gencross})  are dominantly in the most 
attractive relative orientation, $U_\ast =1$, and tend to have equal sizes, 
$\rho_\ast = \bar\rho_\ast$. The remaining equations, determining $(R/\rho)_\ast$ and $\rho_\ast$,  
can be summarized as$^{{\ref{qcd-mod}}}$ 
\begin{eqnarray}
\nonumber
\left(\frac{R}{\rho}\right)_\ast &=& M_W\rho_\ast\,\left( \frac{4\pi M_W/\alpha_W}{\sqrt{\hat s}}\right) 
,\hspace{.3ex} 
\frac{\left( M_W\rho_\ast\right)^2}{2} =
\left[(\xi_\ast -2 )\,
\frac{\partial}{\partial\xi_\ast} \Omega_g (1,\xi_\ast )\right]_{\mid \xi_\ast =2+\left(\frac{R}{\rho}\right)_\ast^2}
    {\rm (QFD)} ,
\\ \label{saddle-coll}
\\ \nonumber
\left(\frac{R}{\rho}\right)_\ast &=& 2\,\frac{Q^\prime}{\sqrt{\hat s}}\,,\hspace{.3ex}  
Q^\prime\,\rho_\ast = \frac{4\pi}{\alpha_s}\,\left[(\xi_\ast -2 )\,
\frac{\partial}{\partial\xi_\ast} \Omega_g (1,\xi_\ast )\right]_{\mid \xi_\ast =2+\left(\frac{R}{\rho}\right)_\ast^2}
\hspace{18.5ex}    {\rm (QCD)},
\end{eqnarray}   
where $(p_1+p_2)^2=\hat s$ denotes the parton-parton center-of-mass (cm) energy. 
To exponential accuracy $(\alpha_g\ll 1)$, the cross-section~(\ref{gencross}) is then given by
\begin{eqnarray}
\label{gen-cross-holy}
\hat\sigma^{(I)} \propto 
{\rm e}^{-\Gamma_\ast} \equiv
{\rm e}^{-\frac{4\pi}{\alpha_g}\,F_g(\epsilon )}
\,,
\end{eqnarray}
where
\begin{eqnarray} 
\epsilon &=& 
               \cases {  
\frac{\sqrt{\hat s}}{4\pi M_W/\alpha_W}
     &{\rm in\ QFD} \,,\cr
\frac{\sqrt{\hat s}}{Q^\prime}
      &{\rm in\ QCD} \,, \cr}
\end{eqnarray}
is a scaled cm energy and 
\begin{eqnarray}
\label{holy-grail}
F_g  &=& 
               \cases {  
\left[1+\Omega_g (1,\xi_\ast )
-(\xi_\ast -2 )\,
\frac{\partial}{\partial\xi_\ast} \Omega_g (1,\xi_\ast )
\right]_{\mid \xi_\ast =2+\left(\frac{R}{\rho}\right)_\ast^2}
     &{\rm in\ QFD} \,,\cr
\left[1+\Omega_g (1,\xi_\ast )\right]_{\mid \xi_\ast =2+\left(\frac{R}{\rho}\right)_\ast^2}
      &{\rm in\ QCD} \,. \cr}
\end{eqnarray}
is the ``holy-grail function~\cite{Mattis:1991bj
}'', determining the fate of instanton-induced hard scattering processes at high energies. 

\FIGURE[t]{
\epsfig{file=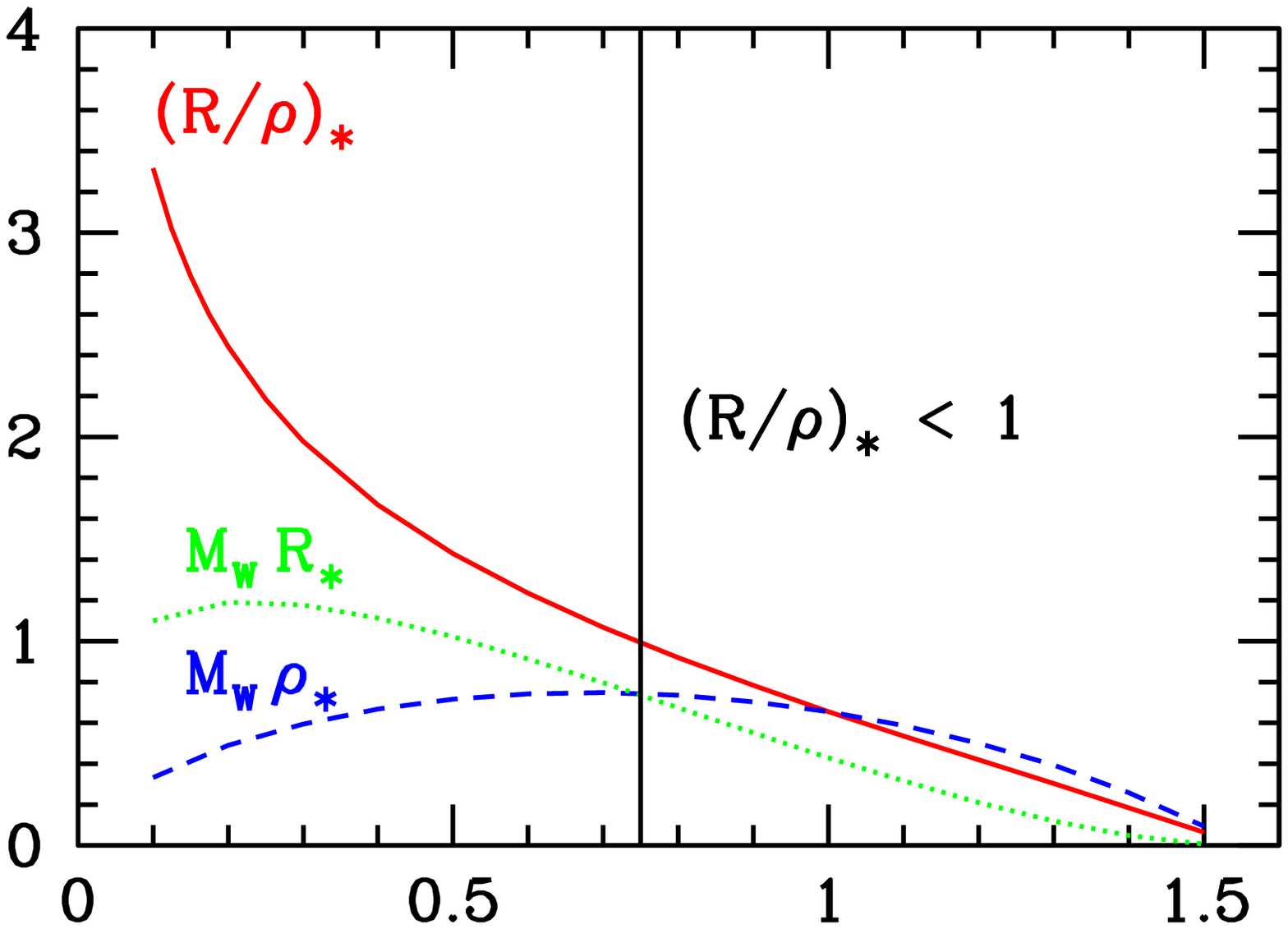,width=7.6cm}%
\mbox{}\vspace{-3cm}
 \epsfig{file=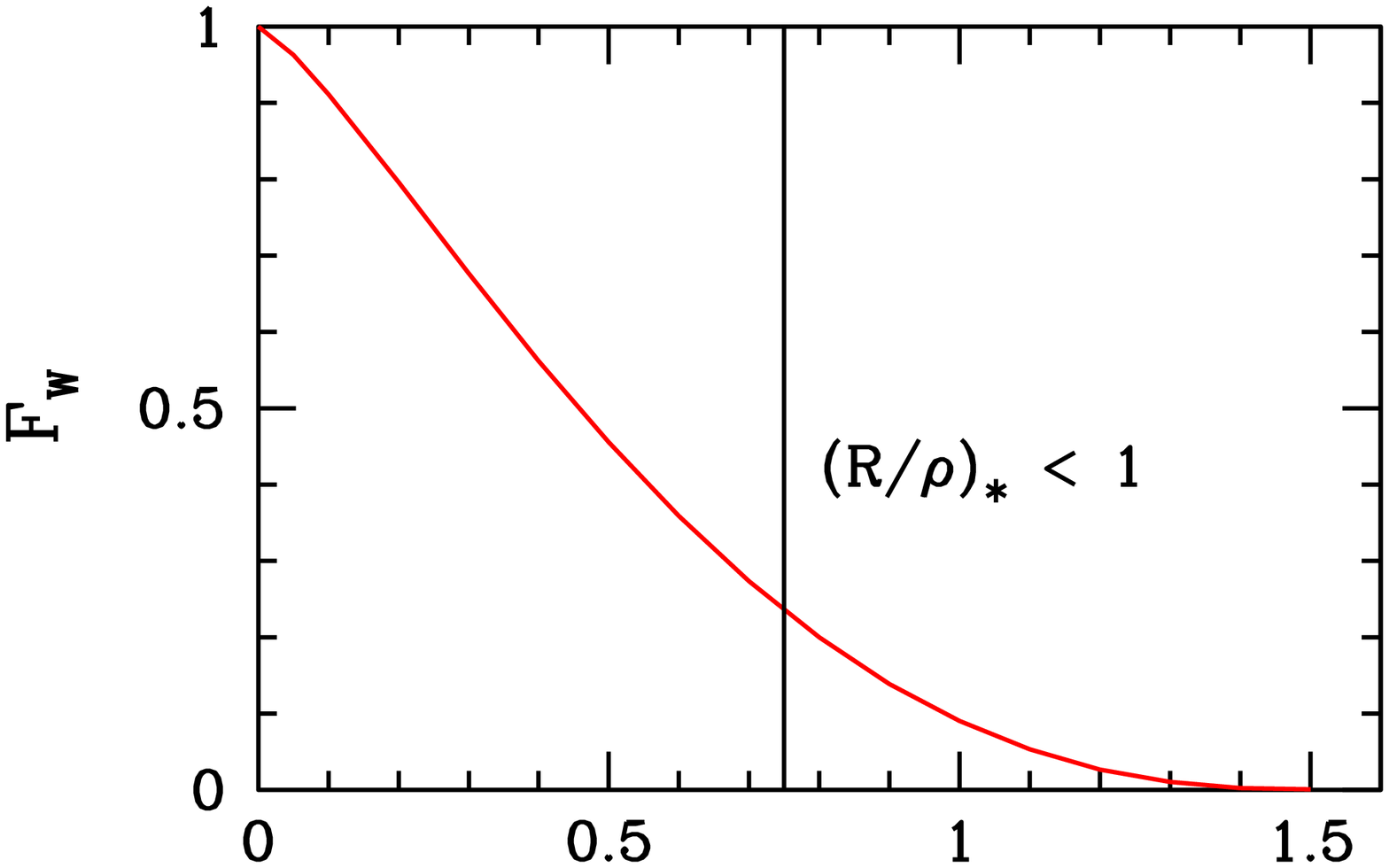,width=7.6cm}%
\mbox{}\vspace{-3cm}
\epsfig{file=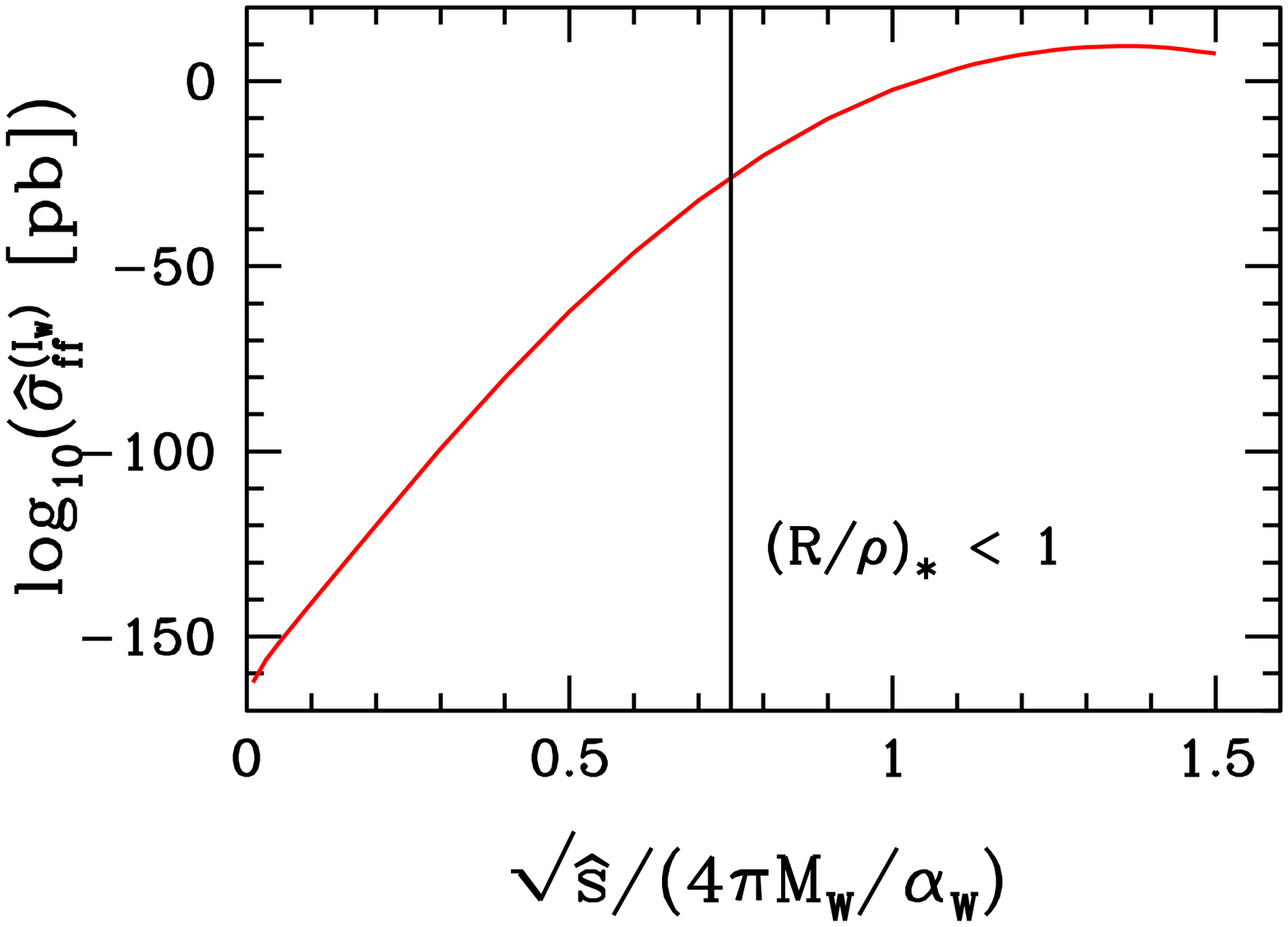,width=7.6cm}%
        \caption{QFD instanton subprocess cross-section related quantities, 
as function of scaled parton-parton center-of-mass energy $\sqrt{\hat s}/(4\pi M_W/\alpha_W )$.
{\em Top:} Saddle point values for collective coordinates~\cite{Khoze:1991mx}. 
{\em Middle:} Holy-grail function, $\hat\sigma^{(I_W)}\propto \exp [-(4\pi/\alpha_W)\,F_W ]$~\cite{Khoze:1991mx}.
{\em Bottom:} Total cross-section $\hat\sigma^{(I_W)}_{\rm ff}$ 
for QFD instanton-induced fermion-fermion scattering, ${\rm f+f}\stackrel{I_W}{\to}{\rm all}$~\cite{Ringwald:2002sw}.   
It stays unobservably small, 
$\hat\sigma_{\rm ff}^{(I_W)}\leq 10^{-26}$~pb, 
in the conservative fiducial
kinematical region corresponding to $(R/\rho)_\ast\geq 1$ 
inferred via the QFD--QCD analogy 
from lattice data and HERA.
If one allows, however, for a slight extrapolation towards
smaller $(R/\rho)_\ast\approx 0.7$, 
the prediction rises to $\hat\sigma_{\rm ff}^{(I_W)}\approx 10^{-6}$~pb.}%
        \label{saddle-holy}}

Both in QFD as well as in QCD, the prediction~(\ref{holy-grail}) for the 
holy-grail function
$F_g(\epsilon )$ decreases monotonically for increasing scaled energy $\epsilon$ from 
its value at zero energy, $F_g(0)=1$. It
approaches zero, $F_g\to 0$, at asymptotic energies, $\epsilon\to\infty$ (cf. Fig.~\ref{saddle-holy} (middle)). 
In other words, at all finite energies, the cross-section~(\ref{gen-cross-holy}) is formally 
exponentially suppressed and there is  no apparent problem with unitarity~\cite{Rubakov:1996vz}.  
It should be noticed, however, that at high cm energies the $I\bar I$-interaction is probed 
at small distances, $(R/\rho)_\ast\sim 1$ (cf. Fig.~\ref{saddle-holy} (top)), making
the semi-classical and saddle-point evaluation unreliable.    
In this connection, the information on the fiducial region in 
$R/\langle\rho\rangle$ of the instanton-perturbative description from QCD lattice simulations 
(cf. Fig.~\ref{qcd-lattice-dist} (right)) can be most appreciated. 
Note furthermore that, in the case of QFD, $M_W R_\ast\,\lwig\, 1$ in the whole
energy range considered in Fig.~\ref{saddle-holy} (top), justifying a posteriori the 
approximation of the full valley interaction in QFD by the one from the pure gauge theory, $\Omega_g$.  

\FIGURE[t]{
\epsfig{file=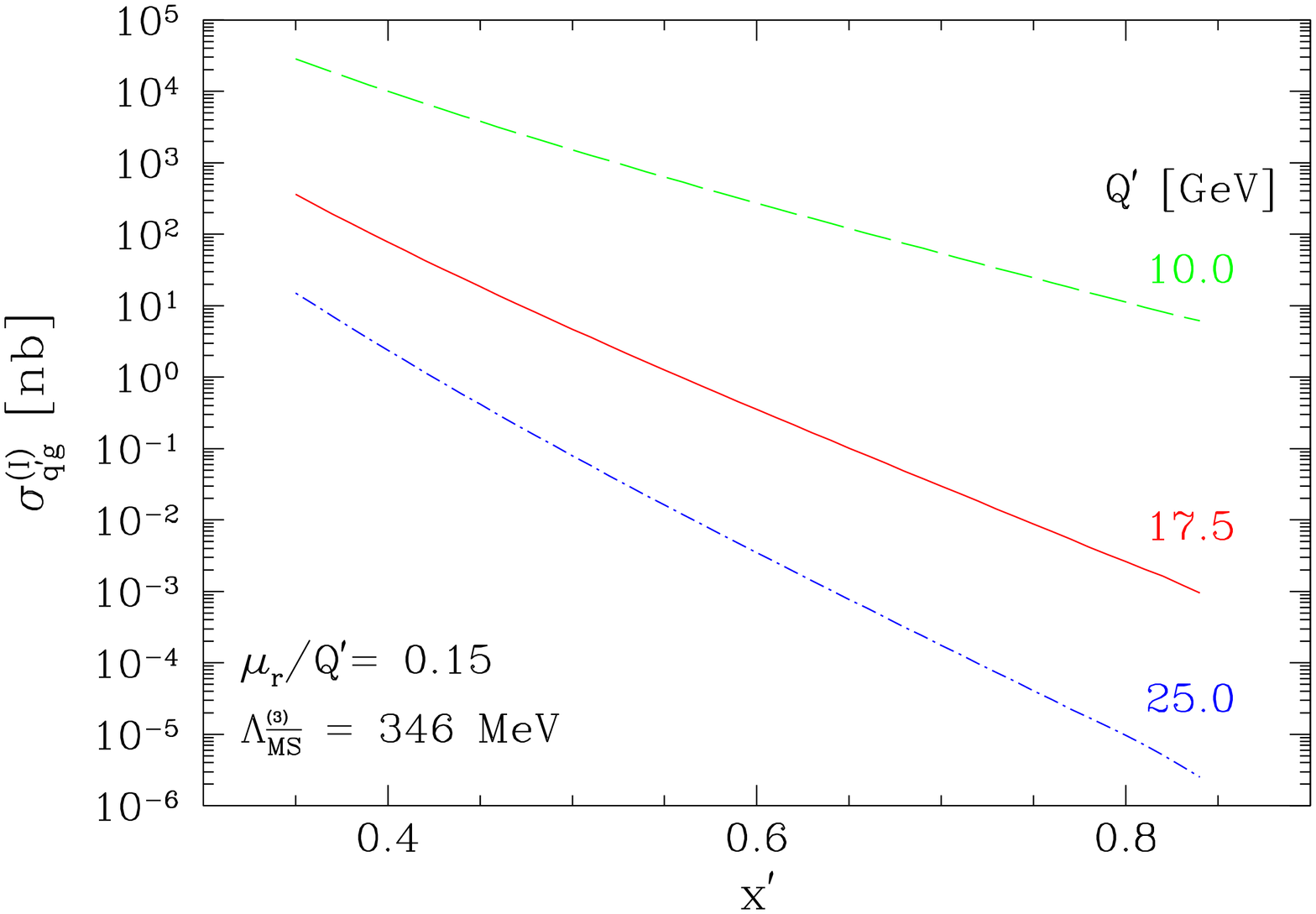,angle=-90,width=8.9cm}%
        \caption{Total cross-section for the QCD instanton quark-gluon initiated 
 subprocess in Fig.~\ref{opt_illu} (right), 
 as a function of the 
 Bjorken variable $x^\prime = Q^{\prime 2}/(Q^{\prime 2}+\hat s)$, 
 for different values of the 
virtuality $Q^\prime$~\cite{Ringwald:1998ek,Ringwald:1999jb}.}%
        \label{isorho}}

The energy dependences of the instanton-induced parton-parton cross sections illustrated 
in Fig.~\ref{saddle-holy} (bottom)  for QFD~\cite{Ringwald:2002sw} and    
in Fig.~\ref{isorho} for QCD in DIS ~\cite{Ringwald:1998ek,Ringwald:1999jb}  
are easily understood on the basis of the 
saddle-point relations above: At increasing energies, smaller and smaller  
$I\bar I$-distances are probed (cf. Fig.~\ref{saddle-holy} (top)), and 
the cross-sections are rapidly growing due to the attractive nature of the $I\bar I$-interaction  in the 
perturbative semi-classical regime (cf. Fig.~\ref{qcd-lattice-dist} (left)).  Furthermore,  in the case of DIS, 
at increasing virtuality $Q^{\prime}$ 
smaller and smaller instantons are probed and the cross-section diminishes  
in accord with the size distribution (cf. Fig.~\ref{size-dist-illu} (right)).

\section{\label{HERA}QCD-instantons at HERA}

Meanwhile, the results of a first dedicated search for QCD instanton-induced processes 
in DIS  have been published by the H1 collaboration~\cite{Adloff:2002ph}. For this search, 
the theory and phenomenology of hard 
QCD instanton-induced processes in DIS developed by Fridger Schrempp and 
myself~\cite{Ringwald:1994kr,Moch:1996bs,Ringwald:1998ek,Ringwald:1999ze,Carli:1998zf,Ringwald:1999jb,Ringwald:2000gt} 
has been used extensively. 
Several observables characterising the hadronic final state of QCD instanton-induced
events (cf. Fig.~\ref{lego}) 
\FIGURE[t]{
\epsfig{file=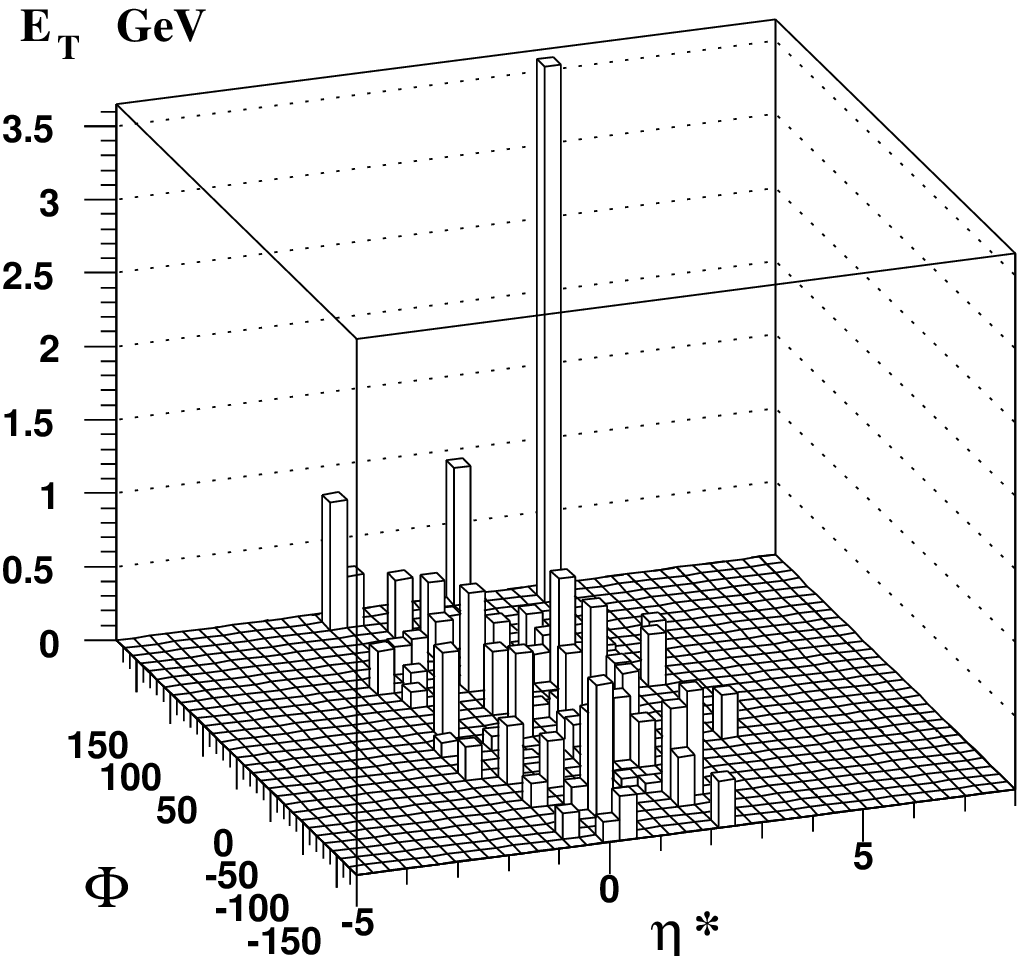,width=8.9cm}%
\caption{Distribution of the transverse energy $E_T$ in the pseudo-rapidity 
$(\eta)$ azimuthal $(\Phi)$ plane in the hadronic cm system 
for a typical QCD instanton-induced event at HERA, generated by QCDINS~\cite{Ringwald:1999jb} 
after typical
detector cuts (from~\cite{Carli:1998zf}). Clearly recognizable are the current jet at 
$\eta \simeq 3, \Phi =160^\textrm{{\small o}}$ and the ``instanton band'' at $0
\stackrel{<}{\sim} \eta \stackrel{<}{\sim} 2$, the latter reflecting the isotropic multi-particle 
nature of an instanton-induced final state.} 
   \label{lego}}
were exploited to identify a potentially instanton-enriched domain. 
The results obtained are intriguing but non-conclusive. An excess of events with instanton-like topology
over the expectation of the standard DIS background is observed,  which, moreover,  is compatible 
with the instanton signal (cf. Fig.~\ref{H1-observ}). 
After combinatorial cuts on instanton-sensitive observables, which suppress 
the background by about a factor of ${\mathcal O}(10^3)$, $484$ events are found in the data, while
the standard DIS background Monte Carlo simulations MEPS and CDM predict  $304^{+21}_{-25}$ and  
$443^{+29}_{-35}$, respectively\footnote{The quoted errors on the expected event numbers include 
the statistical and the experimental systematic uncertainties. For details about MEPS and CDM, see 
Ref.~\cite{Adloff:2002ph}}. 
\FIGURE[t]{
\epsfig{file=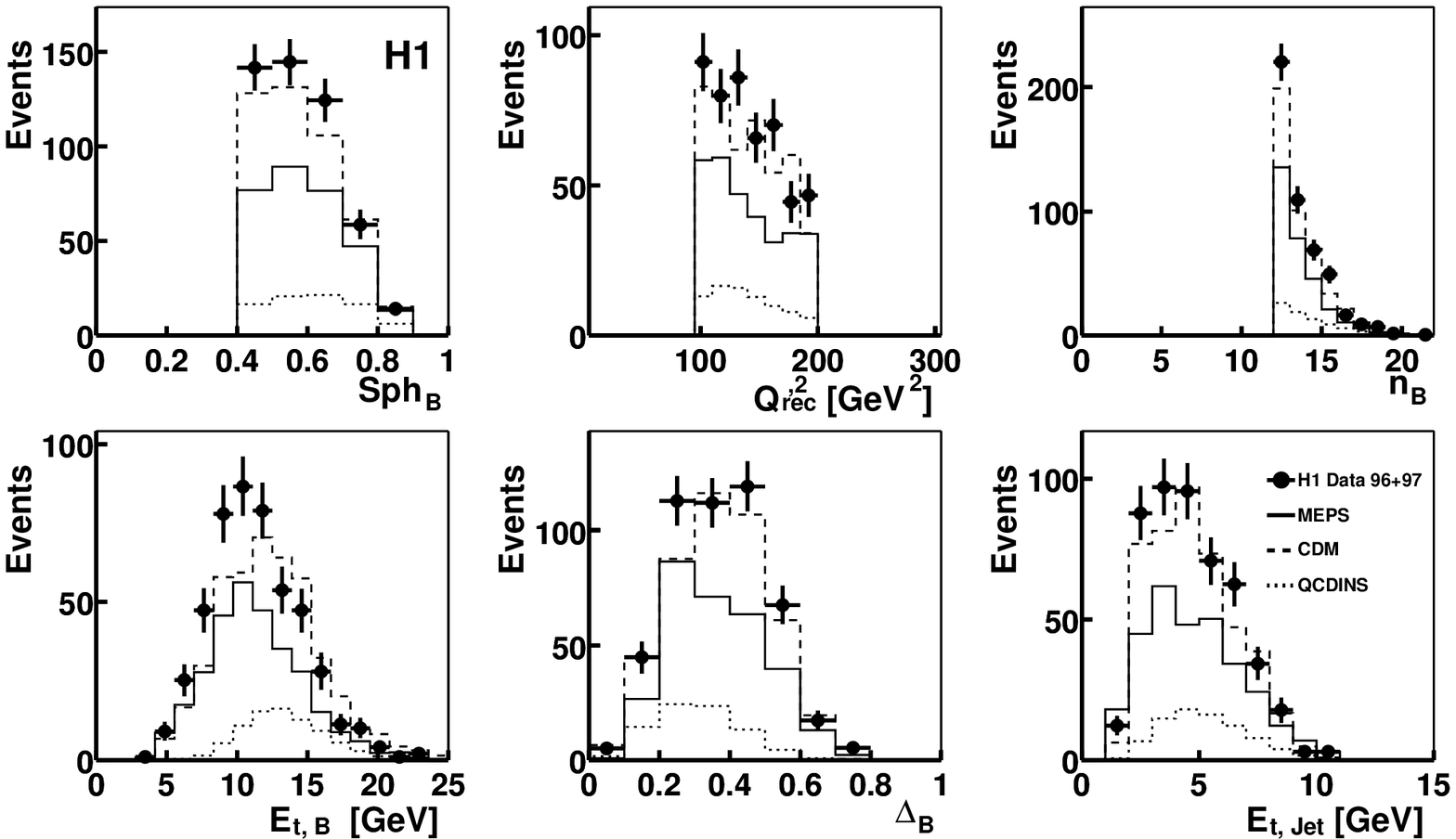,width=14cm}%
        \caption{Distributions of DIS final-state observables (cf. Fig.~\ref{lego}) 
after combinatorial cuts, as measured by the H1 collaboration at HERA~\cite{Adloff:2002ph}. 
{\em Upper panel from left to right:} 
{\em a)} Sphericity in the instanton band, {\em b)} reconstructed virtuality, 
{\em c)} charged particle multiplicity in the instanton band.
{\em Lower panel from left to right:} 
{\em d)} transverse energy in the instanton band, 
{\em e)} isotropy variable, {\em f)} transverse energy of current jet.
Data (filled circles), the predictions from the standard DIS Monte Carlo simulations 
MEPS and CDM (solid and dashed line, respectively) and from the QCD instanton
Monte Carlo simulation QCDINS~\cite{Ringwald:1999jb} (dotted) are shown. 
The data exhibit an excess over the 
predictions from MEPS and CDM, which is compatibel with the instanton signal.
In view of the uncertainties of the prediction of the standard DIS background, the excess 
can not be claimed to be significant, however.}%
        \label{H1-observ}}
However, the observed excess can not be claimed to be significant
given the uncertainty of the Monte Carlo simulations of the standard DIS background.
Therefore, only upper limits on the cross-section for QCD instanton-induced processes
are set, dependent on the kinematic domain considered~\cite{Adloff:2002ph}. From this 
analysis one may infer, via the saddle point correspondence, 
that the cross-section calculated within instanton-perturbation theory 
is ruled out for $(R/\rho)_\ast\lwig 0.84$, in a range $0.31\ {\rm fm}\,\lwig \rho_\ast\lwig\,0.33$ fm 
of effective instanton sizes~\cite{Ringwald:2002sw}. 
It should be kept in mind, however, that 
in the corresponding -- with present statistics accessible -- kinematical range 
the running coupling is quite large, $\alpha_s(\rho^{-1}_\ast)\approx 0.4$, 
and one is therefore not 
very sensitive\footnote{This is of course welcome for the QCD instanton searches at HERA, because 
it makes predictions  for the bulk of data quite reliable.} to the 
$I\bar I$-interaction, 
which appears in the exponent with coefficient $4\pi/\alpha_s\approx 31$. Instanton-induced 
rates in QFD, on the other hand, are extremely sensitive to $\Omega$, since $4\pi/\alpha_W\approx 372$.       
An extension of the present H1 limit on $(R/\rho)_\ast$ towards smaller $\rho_\ast$ and  
$\alpha_s(\rho_\ast^{-1})$, which should
be possible with increased statistics at HERA II, would be very welcome.
At present, the data do not exclude the cross-section predicted by instanton-perturbation theory 
for small $(R/\rho )_\ast \gwig 0.5$, as long as one probes only very small 
instanton-sizes $\rho_\ast\ll 0.3$~fm.

\section{\label{VLHC}QFD-instantons at VLHC?}

Let us finally discuss the result of a state of the art evaluation of the cross-section~(\ref{gencross})
for QFD~\cite{Ringwald:2002sw}, including all the prefactors -- an analogous evaluation has been presented for QCD
in DIS in Ref.~\cite{Ringwald:1998ek}. 
The prediction\footnote{\label{guess}At $\epsilon\sim 1$ it should be rather called an 
educated extrapolation or guess.} 
for the QFD instanton-induced fermion-fermion cross-section -- as relevant at VLHC at the parton level --  is 
displayed in Fig.~\ref{saddle-holy} (bottom) as a function of the scaled fermion-fermion cm energy
$\epsilon =\sqrt{\hat s}/(4\pi M_W/\alpha_W)$, for a choice $\mu = M_W$ 
of the renormalization scale. 
In the strict region of instanton-perturbation theory, 
$\epsilon\ll 1$, 
the cross-section is really tiny, e.g. $\hat\sigma_{\rm ff}^{(I_W)}\approx 10^{-141}$~pb at 
$\epsilon\approx 0.1$, but steeply growing. 
Nevertheless, it stays unobservably small, 
$\hat\sigma_{\rm ff}^{(I_W)}\lwig 10^{-26}$~pb for $\epsilon\,\lwig\, 0.75$, 
in the conservative fiducial
kinematical region corresponding to $(R/\rho)_\ast\gwig 1$ 
inferred via the QFD--QCD analogy 
from lattice data and HERA. If one makes, however, a slight extrapolation towards
smaller $(R/\rho)_\ast\approx 0.7$ -- still compatible with lattice results and HERA -- 
the prediction$^{\ref{guess}}$ rises to $\hat\sigma_{\rm ff}^{(I_W)}\approx 10^{-6}$~pb at $\epsilon\approx 1$, 
corresponding to a parton-parton cm energy 
of about $30$~TeV. In this case, QFD instanton-induced $B+L$ violating events will have 
observable rates at VLHC, which has a projected proton-proton cm energy of $\sqrt{s}=200$~TeV  
and a luminosity of about ${\mathcal L}\approx 6\cdot 10^5$~pb$^{-1}$\,yr$^{-1}$~\cite{vlhc}.   
An exciting phenomenology at VLHC will emerge if this possibility is realized in 
nature~\cite{Farrar:1990vb,
Gibbs:1994cw
}. The high transverse energy of the final state, combined with the large number
of high $p_T$ leptons and the inability to resolve jets means that any conventional 
Standard Model background can be easily separated. While the prospects to directly verify 
$B$ violation are poor, the verification of $L$ violation seems to be possible 
if one succeeds in collecting a sample of $10^{2\div 3}$ 
QFD instanton-induced events~\cite{Gibbs:1994cw
}.   

Let us finally speculate about the possibility that 
the prediction$^{\ref{guess}}$ in Fig.~\ref{saddle-holy} (bottom) is valid even at higher energies, 
corresponding to even smaller $(R/\rho)_\ast$. In this case, we can expect to be able to see the first signs 
of electroweak sphaleron production in present day or near future cosmic ray facilities and 
neutrino telescopes~\cite{Morris:1991bb
}, even before the commissioning of VLHC. 

In the meantime, we have a lot of opportunities to improve our 
understanding of QCD instantons on the lattice and in deep-inelastic scattering at HERA, 
with important implications also for QFD instantons at very high energies.

\acknowledgments

I would like to thank my colleague F.~Schrempp for the nice long-term collaboration on 
instanton-induced scattering processes in the Standard Model and a careful reading
of the manuscript.  
Furthermore, I would like to thank the organizers of 
this workshop 
for inspiration and encouragement 
of this review.  


\end{document}